\documentclass[iop]{emulateapj}
\usepackage{natbib}
\shorttitle{Cluster Halo Shape Properties}
\shortauthors{Groener et al.}

\begin{document}
\title{Shape Profiles and Orientation Bias for \\
Weak and Strong Lensing Cluster Halos} 

\author{A. M. Groener and D. M. Goldberg}
\affil{Physics Department, Drexel University,
    Philadelphia, PA 19104}
\email{Austen.M.Groener@drexel.edu}

\begin{abstract}
  We study the intrinsic shape and alignment of isodensities of galaxy
  cluster halos extracted from the MultiDark MDR1 cosmological
  simulation. We find that the simulated halos, are extremely prolate
  on small scales, and increasingly spherical on larger ones. Due to
  this trend, analytical projection along the line of sight produces an
  overestimate of the concentration index as a decreasing function of
  radius, which we quantify by using both the intrinsic distribution
  of 3D concentrations ($c_{200}$) and isodensity shape on weak and
  strong lensing scales. We find this difference to be $\sim 18\%$
  ($\sim 9\%$) for low (medium) mass cluster halos with intrinsically
  low concentrations ($c_{200}=1-3$), while we find virtually no
  difference for halos with intrinsically high
  concentrations. Isodensities are found to be fairly well-aligned
  throughout the entirety of the radial scale of each halo
  population. However, major axes of individual halos have 
  been found to deviate by as much as $\sim 30^{\circ}$. We also
  present a value-added catalog of our analysis results, which we have
  made publicly available to download. 
\end{abstract}

\keywords{galaxies: clusters: general -- cosmology: dark matter -- gravitational
lensing: strong -- gravitational lensing: weak }

\section{Introduction}

Galaxy clusters represent the most massive of the virialized structures in
the universe, whose development comes after billions of years of
hierarchical merging of matter. Clusters offer us key insights
into the structure formation process as predicted by the standard
Lambda Cold Dark Matter ($\Lambda$CDM) cosmological paradigm
\citep{CorlessKing2009,EttoriEtAl2009}. Radial 
density profiles, mass functions, and the baryon mass fraction are all
used to constrain cosmological parameters
\citep{NavarroEtAl2004,Voit2005,DiaferioEtAl2008,AllenEtAl2011}, and rely 
on the accurate measurement of the distribution of mass within
clusters, total mass, and proper characterization of halo substructure.  

Gravitational lensing has proven an incredibly useful tool in
creating detailed maps of projected density without requiring any
assumptions regarding either a halo's dynamical state or hydrostatic
equilibrium of its gas. Recently, strong and weak 
gravitational lensing methods have been used side by side to constrain
the distribution of mass on different scales within the same cluster
lens \citep[for Abell 1689 for example, see][]{BroadhurstEtAl2005,HalkolaEtAl2006,DebEtAl2012}.    

Observations suggest that cluster halos exhibit triaxial geometry (see
\citealt{LimousinEtAl2013} for a general discussion) in
the optical distribution of light \citep{CarterMetcalfe1980,Binggeli1982} in X-ray
\citep{FabricantEtAl1985,LauEtAl2012} and in weak
\citep{EvansBridle2009,OguriEtAl2010,OguriEtAl2012} and strong 
lensing \citep{SoucailEtAl1987}. Observational evidence for
triaxiality is matched by theory in large scale N-body simulations of
structure formation, exhibiting a preference for prolateness over
oblateness
\citep{FrenkEtAl1988,DubinskiCarlberg1991,WarrenEtAl1992,ColeLacey1996,JS2002,HopkinsEtAl2005,BailinSteinmetz2005,KasunEvrard2005,PazEtAl2006,AllgoodEtAl2006,BettEtAl2007,MunozEtAl2011,GaoEtAl2012}.       

However, the over-simplification of the spherical halo approximation
has serious consequences on their utility as cosmological probes. Of
particular importance to cosmology is the concentration of the halo,
$c_{200}$, a measure of how centrally concentrated the halo
is. Qualitatively, the concentration describes the steepness of the
inner density profile of the halo, where larger concentrations give
rise to a steeper (``cuspier'') inner density profile, and conversely,
smaller concentrations produce a more well-defined central core,
indicating a shallower inner density profile. It has been shown
through simulations that the concentration scales inversely with halo
mass in the form of a power-law, and has also been more intuitively
related to the formation time of a certain fraction of the mass of the
main halo progenitor \citep{GiocoliEtAl2012a}.    

A somewhat long-standing discrepancy between observations and what
$\Lambda$CDM predicts is what is known as the over-concentration
problem, wherein observed cluster concentrations are nearly twice what
simulations predict \citep{BroadhurstEtAl2008,OguriEtAl2009}. 
\citet{BaheEtAl2012} have shown that when observed
in near alignment with their major axes, weak lensing reconstructed 
concentrations are systematically larger by up to a factor of 2 for
Millennium Simulation cluster halos. Similarly, \citet{OguriBlandford2009}
find in their semi-analytic study that the most massive triaxial halos
($\sim 10^{15} \mathrm{h}^{-1 } \mathrm{M}_{\odot}$) produce the largest Einstein radii if
they're viewed preferentially along their major axes, thereby
increasing their effectiveness as strong gravitational lenses. Through
ray-tracing of high-resolution N-body simulations,
\citet{HennawiEtAl2007} have shown that strong lensing clusters tend
to have their principle axes aligned along the line of sight.  
The projection bias due to triaxial halos associated with lensing
methods has been fairly successful in describing much of the discrepancy
between predictions and measurements. \citet{GiocoliEtAl2012b} have
shown that knowledge of the elongation along the line of sight can
help in correcting mass estimates, though a small negative bias
remains in concentration.

Additional complications arise in this picture of cluster halos in
that halo axis ratios change as a function of
radius. \citet{FrenkEtAl1988} and \citet{ColeLacey1996} found that
simulation halos become more spherical towards the center, whereas
\citet{DubinskiCarlberg1991}, \citet{WarrenEtAl1992}, and
\citet{JS2002} have concluded the opposite. Nearly all are in agreement
that there is good alignment between isodensity (or isopotential)
surfaces on most scales. More recent work done by
\citet{HayashiEtAl2007} show that axis ratios of both isopotential and
isodensity surfaces consistently increase with radius for seven
galaxy-scale simulations.   

Though lensing concentrations are systematically higher than their
predicted values simply due to orientation bias, there
are other reasons to believe that clusters identified by their strong
lensing features are a biased population. For one, the most massive
clusters are simply more effective gravitational lenses,
preferentially sampling the highest region of the cluster mass
hierarchy \citep{CN2007}.  

Many studies have employed joint weak and strong lensing reconstruction
techniques in order to probe much larger regions of the radial density profile.
However, weak and strong lensing can sometimes independently produce
vastly different results, and are rather sensitive to the a priori
assumptions made regarding the distribution of mass within the cluster
lens. \citet{BroadhurstEtAl2005} and
\citet{HalkolaEtAl2006} both find weak lensing concentrations to be
much larger than ones produced by strong lensing of the well-known
cluster Abell 1689, when a spherical model is used. Weak + strong
lensing analyses of Abell 1689 have generally been in agreement with
one another, however, concentrations remain inconsistent with what
theory predicts
\citep{Clowe2003,HalkolaEtAl2006,LimousinEtAl2007}. Only when a 
triaxial halo model is employed do theory 
and observation come into agreement (\citealt{OguriEtAl2005}; for a
complete overview of the cluster Abell 1689, see $\S$5 of
\citealt{LimousinEtAl2013}). The model assumptions and priors used in
weak and strong lensing methods can produce large uncertainty in
reconstructed parameters. However, physical features of the cluster
halos, for example the combination of an orientation bias together with an
intrinsic trend in halo shape as a function of radius could perhaps
also lay at the heart of discrepancies of this nature, and work to
diminish the accuracy of lensing techniques as stand-alone tools as
well as joint techniques. 

Lastly, ongoing baryonic physics within galaxy clusters
(specifically cooling, star formation, and AGN feedback) has the
potential to significantly alter the distribution of mass within
clusters, and is absent from many simulations of structure
formation. The lensing cross-section depends sensitively on the
addition of baryons to cluster simulations, and can be boosted by a
factor of a few
\citep{PuchweinEtAl2005,WambsganssEtAl2008,RozoEtAl2008}. 
Studies which do not include AGN feedback suffer from over-cooling,
and the addition of this component reduces the enhancement of the
lensing cross-section to at most a factor of two
\citep{MeadEtAl2010}. \citet{KilledarEtAl2012} find Einstein radii of
clusters to be larger only by $5\%$ when AGN feedback is included for
$z_{s} = 2$ (increasing to $10-20\%$ for lower source redshifts).

In this paper we present a study of the shape and alignment of
isodensity surfaces from simulated cluster halos of the MDR1
cosmological simulation \citep{PradaEtAl2012} throughout a range of radial
scales. Differences between the concentration parameter on weak and
strong lensing scales will be quantified by the analytical projection
of NFW halos, for the specific case that their major axes point along
our line of sight. This paper is organized as follows. In section
$\S$2, we discuss the extension of the spherical NFW model and present
prolate spheroidal simplifications of halo properties upon projection
of the 3D NFW profile along the line of sight. In $\S$3 we define our
samples and methods we will use to analyze halo intrinsic
properties. In $\S$4 we present our findings on weak and strong
lensing scales within clusters, and discuss our results in $\S$5. And
lastly, in $\S6$ we add a general discussion of future work.  

\section{Projections of Triaxial NFW Halos}
The Navarro-Frenk-White density profile \citep{NFW1996} has been shown
to describe simulation halos over many decades of mass. The model in
its simplest form contains two free parameters, $r_{s}$ and $c_{200}$
(implicitly in $\delta_{c_{200}}$; Equation 3): 
\begin{equation}
\rho_{\mathrm{NFW}}(r) = \frac{\rho_{cr} \delta_{c_{200}}}{r/r_{s}(1+r/r_{s})^{2}}
\end{equation}
The first parameter is the scale radius $r_{s}$, which describes the
turning point of the profile - the point at which the slope of the
logarithmic density is equal to -2. The second is the
concentration, which we define as 
\begin{equation}
c_{200} \equiv r_{200}/r_{s}
\end{equation}
where $r_{200}$ is the radius at which the average density contained
within it is a factor of 200 times larger than the critical density of the
universe at the redshift of the halo. The characteristic overdensity
of the cluster is defined in terms of the concentration in the
following way: 
\begin{equation}
\delta_{c_{200}} = \frac{200}{3} \frac{c_{200}^{3}}{\ln{(1+c_{200})} - \frac{c_{200}}{(1+c_{200})}}
\end{equation}
Mass can now be defined in terms of $r_{200}$ and the
cosmology-dependent critical density of the universe at the redshift
of the halo. 
\begin{equation}
M_{200}(z) = \frac{4}{3} \pi r_{200}^{3} \cdot 200
\rho_{cr}(z)
\end{equation} 

The concentration is an important measurement, since it has been
shown through simulations to scale inversely (though somewhat weakly)
with the virial mass of the halo
\citep{NFW1996,BullockEtAl2001,HennawiEtAl2007}. More recent work done
by \citet{PradaEtAl2012} has actually revealed a new flattening/upturn
feature in the concentration-mass relationship for high redshift high
mass halos in the Millennium-I and II, Bolshoi, and
MultiDark simulations. 

Simulations show that halos deviate from spherical symmetry by
a considerable amount, with a preference for prolate over oblate
spheroids. \citet{Doroshkevich1970} has made the case that triaxial
collapse is a necessary outcome of structure formation models which
are seeded by Gaussian random initial conditions. The spherical NFW
profile can be extended to accommodate triaxial halos by redefining
the radial coordinate as an ellipsoidal coordinate:  
\begin{equation}
\zeta^{2} = \frac{x^{2}}{c^{2}} + \frac{y^{2}}{b^{2}} + \frac{z^{2}}{a^{2}}
\end{equation}
where $a,b,c$ are the semi-major, -intermediate, and -minor axes of
the ellipsoidal shell under consideration. Each set of semi-axes is
unique to a specific radial value within each halo by the relationship
$abc = r^{3}$.   

The projection of an arbitrary triaxial halo onto a 2-dimensional plane is
a unique process. However, the reverse process is degenerate. To
understand the limits of projection biases due to orientation, we 
simplify the generalized projection of triaxial halos found in
\citet{SerenoEtAl2010b} assuming a prolate spheroidal geometry. The
following observed (projected) parameters can be expressed in terms of
the angle $\theta$ between the line-of-sight and 
the major axis of a prolate spheroid, as well as a single axial ratio
intrinsic to the cluster (Table 1).

\begin{table}
\caption{Prolate Spheroidal Geometry}
\label{table-1}
\begin{tabular}{cc}
\hline
\hline
\textbf{Projected Quantity} & \textbf{Prolate Spheroidal Expression} \\ 
\hline 
$Q$ &  $\frac{q}{\sqrt{\sin^{2}{\theta} + q^{2}\cos^{2}{\theta}}}$\\
\hline
$R_{s}$ &  $\frac{q}{Q}  r_{s}$\\
\hline
$\delta_{C}$ & $\frac{Q^{2}}{q} \delta_{c}$\\
\hline
\end{tabular}

\medskip
Shown here are the analytical expressions for projected quantities
expressed  in terms of intrinsic quantities and halo orientation used
throughout this study. As a general rule, projected quantities will be 
capitalized while intrinsic quantities will not be.
\end{table}

\section{Sample and Methods}
\subsection{Simulation Sample}

We aim to quantify the effect which line-of-sight alignment of
triaxial cluster halos has upon projected concentrations on both
weak and strong lensing scales in the presence of slowly evolving
isodensity shapes. We study clusters from the MDR1 cosmological
simulation \citep{PradaEtAl2012} of the MultiDark Project\footnotemark
\footnotetext[1]{http://www.multidark.org/MultiDark/}.  MDR1 is a 
dark matter only simulation which uses $2048^{3}$ particles in
a box 1 h$^{-1}$Gpc on a side. The mass of simulation particles is $8.721 \times 10^{9}$ 
h$^{-1}$M$_{\odot}$, with a resolution of 7 h$^{-1}$kpc. The
simulation uses results from WMAP5 as its cosmology and was run with
the Adaptive-Refinement Tree (ART) code \citep{Kravtsov1997}.

\begin{figure}
\centering
\includegraphics[width=0.5\textwidth]{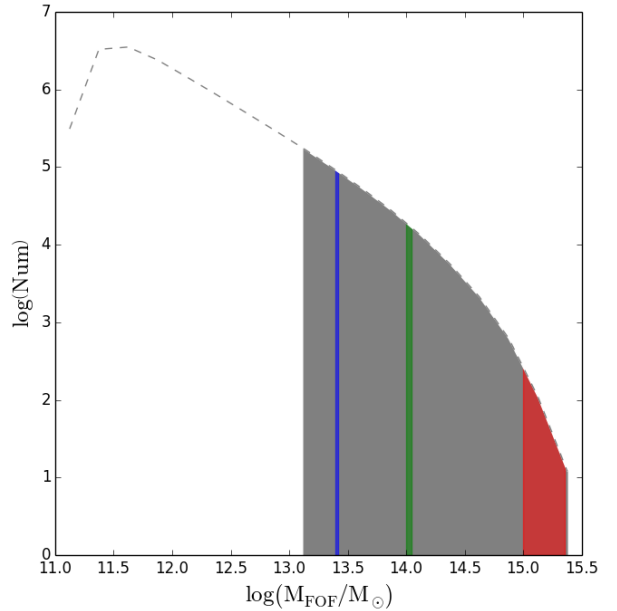}      
\caption{Shown in black is the MDR1 mass function using the FOF
  algorithm. The gray shaded region shows the cluster halo regime. Overplotted are the low (blue), medium
  (green), and high (red) mass samples extracted from the database.} 
\label{fig:MDR1 Mass Function}
\end{figure}

The MDR1 simulation database \citep{Riebe2013} contains halo catalogs which are
found using two different halo finding algorithms (``Bound Density
Maximum'' BDM, and  ``Friends-of-Friends'' FOF) taken at 85 redshift
snapshots. The masses of halos found through these two methods range
from $1.7 \times 10^{11} \mathrm{h}^{-1} \mathrm{M}_{\odot}$ - $1.6 \times 10^{15}
\mathrm{h}^{-1} \mathrm{M}_{\odot}$. We impose a lower cutoff in halo mass of $1 \times 10^{13}
\mathrm{h}^{-1} \mathrm{M}_{\odot}$ to mark the beginning of the cluster halo regime which
are considered for this study. Figure 1 shows the mass function of the
sample at redshift $z=0$.   

In our study, three halo samples were extracted from the MDR1 FOF
table (relative linking length of 0.17) of the MultiDark Database for
further study. Smaller linking lengths (e.g. halo substructures) are
available, however these were purposefully left out since it is beyond
the scope of this study.
\begin{itemize}
\item Low Mass: $2.5 - 2.6 \times 10^{13} \mathrm{h}^{-1} \mathrm{M}_{\odot}$ [6007 halos]
\item Medium Mass: $1.0 - 1.1 \times 10^{14} \mathrm{h}^{-1} \mathrm{M}_{\odot}$ [2905 halos] 
\item High Mass: $> 1.0 \times 10^{15} \mathrm{h}^{-1} \mathrm{M}_{\odot}$ [121 halos]
\end{itemize}

It should be noted here that our results are derived using one
particular agglomerative, single-linkage clustering algorithm
(FOF). Studies have been conducted which compare various halo-finding
algorithms for simulation data (see \citet{KnebeEtAl2011} for a
general review of such algorithms and how they perform in identifying
structures from simulations). \citet{DespaliEtAl2013} have shown that
halo shape can depend upon the choice of method used to identify halos
from the overall simulation volume.

\subsection{Methods}
Following, for example, \citep{WarrenEtAl1992,ShawEtAl2006}, we
compute the moment of inertia tensor, from which we obtain the average
shape of the halo at a given radius:
\begin{equation}
I_{ij} = \sum_{n=1}^{n=N} m_{p} (r_{i,n} - \bar{r_{i}})(r_{j,n} - \bar{r_{j}})
\end{equation}
where $r_{i,n}$ is the coordinate of the $n^{th}$ particle in the
$i^{th}$ direction (where $i,j \in {x,y,z}$), and $m_{p}$ is the mass
of the simulation particle. Finding the eigenvalues
and eigenvectors of $I$ can uniquely determine the orientation and
axis ratios. 

This method of determining shape has its
drawbacks. \citet{ShawEtAl2006}, \citet{JingSuto2002}, and
\citet{BailinSteinmetz2004} have all found that this approach often
fails to converge in high resolution simulations with substantial
substructure. Additionally, substructures of fixed mass will affect
the components of the moment of inertia tensor on larger 
scales than it will on small scales. In order to correct for this, we apply a
Gaussian weighting function, $w_{g} (\zeta)$, (to Equation 6) which
matches both the shape and orientation of each bounding ellipsoid
within the iterative process. 

 We find good convergence with this method for MDR1 halos. For our
 purposes, calculating the shape of each halo using the inertia
 tensor, with the addition of a triaxial, Gaussian weighting function,
 will be more than adequate to describe the macrostructure of the
 parent halo.   

\subsection{Non-Virialized Halos}
Mergers are common physical processes in the
formation of galaxy clusters
\citep{PS1974,BondEtAl1991,LaceyCole1993}. Special care is taken to 
separate out halos which are better fit by more than one halo for each
mass sample, since parameters like concentration and scale radius are
only properly defined for a single halo profile. The Mean Shift
clustering algorithm \citep{FukunagaHostetler1975} is used on low and
medium mass clusters, while the K-Means \citep{HartiganWong1979}
clustering algorithm is used for high mass halos.  

The Mean Shift algorithm is a non-parametric algorithm which 
requires no initial guess for the number of clusters, making it
ideal for handling clusters of arbitrary shape or number. It locates
local density maxima and uses a tuning parameter to associate each particle's
membership to a corresponding maximum. Conversely, the K-Means
algorithm necessarily requires k-clusters to group particles into. K-Means scales as $O(knT)$ (where $k$ is
the number of clusters, $n$ is the number of points, and $T$ is the
number of iterations) and therefore was chosen for the high mass sample of 121 halos.

Additionally, to prevent contamination of each sample by unrelaxed or
actively merging structures which may be well-fit by a single model,
halos are checked for virialization using the following method
\citep{ShawEtAl2006}:   
\begin{equation}
\beta = \frac{2T_{0}-E_{s}}{W_{0}} + 1
\end{equation}
where $T_{0}$ is the total kinetic energy, $W_{0}$ is the total
potential energy, and $E_{s}$ comes from the pressure of the outer
perimeter of the halo, an important contribution which comes from the
fact that cluster halos are not isolated systems. A cut is made at
$\beta > -0.2$ in order to remove halos which have sufficient pressure
at their virial radii, indicating that they are currently in a state
of collapse \citep{ShawEtAl2006}. 

All analysis results for each cluster halo within this study have been
stored in a downloadable database\footnotemark
\footnotetext[2]{http://www.physics.drexel.edu/$\sim$groenera/zodb$\_$file.fs}. 

\section{Results}
Based upon the above criteria that a halo must be both virialized and 
best fit by a single component halo model, we find that  $55.3\%$,
$42.0\%$, and $17.4 \%$ of halos in the low, medium, and high mass
samples make it through our selection criteria. Additionally, nearly
all of these cluster halos can be described as being prolate
ellipsoids, becoming marginally more spherical with increasing
distance from the cluster center (See Figure 2; See Table 2 for a
summary of these shape results).  

\begin{table*}
\begin{center}
\caption{Cluster Halo Geometry}
\label{table-2}
\begin{tabular}{@{}lcccccc}
\hline
\hline
\textbf{Radial Scale}& & & & \textbf{Low} & \textbf{Medium} & \textbf{High} \\ 
\hline
$0.5 \cdot r_{200}$ & & $\bar{p} \pm \sigma_{p}$ & & $0.67 \pm 0.15$ & $0.62 \pm 0.14$ & $0.56 \pm 0.13$ \\
& & $\bar{q} \pm \sigma_{q}$ & & $0.53 \pm 0.12 $ & $0.49 \pm 0.11$ & $0.42 \pm 0.06$ \\
\hline
$r_{200}$ & & $\bar{p} \pm \sigma_{p}$ & & $0.71 \pm 0.13$ & $0.66 \pm 0.13$ & $0.54 \pm 0.15$ \\
& & $\bar{q} \pm \sigma_{q}$ & & $0.57 \pm 0.11$ & $0.52 \pm 0.10$ & $0.42 \pm 0.08$ \\
\hline
$2 \cdot r_{200}$ & & $\bar{p} \pm \sigma_{p}$ & & $0.69 \pm 0.12$ & $0.67 \pm 0.12$ & $0.54 \pm 0.12$ \\
& & $\bar{q} \pm \sigma_{q}$ & & $0.55 \pm 0.10$ & $0.53 \pm 0.10$ & $0.43 \pm 0.08 $ \\
\hline
\end{tabular}

\medskip
Reported here are the sample mean (standard deviation) values of the semi-intermediate to semi-major axis ratio
$p$, and semi-minor to semi-major axis ratio $q$ for each mass sample (virialized, non-mergers) at various physical scales of interest.
\end{center}
\end{table*}

\begin{figure}
\begin{center}$
\begin{array}{c}
\includegraphics[scale=0.4]{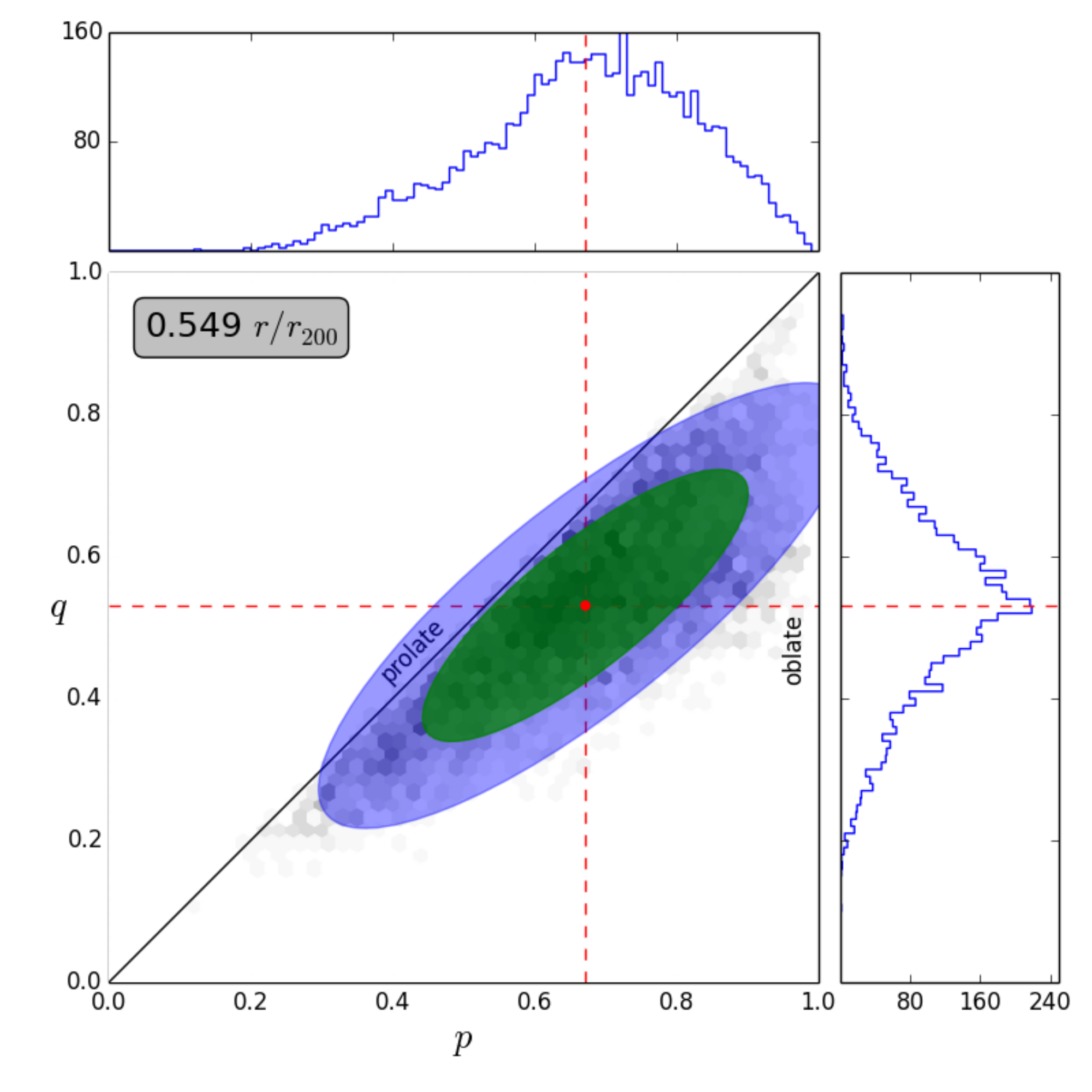}
\end{array}$
\end{center}
\caption{Shown above is the distribution of isodensity shapes for the low mass halo sample at a radial scale of $\sim 0.5 r_{200}$. The green and blue shaded regions are the 1- and
  2-$\sigma$ gaussian error ellipses, and red indicates the sample mean.}
\end{figure}

Intrinsic NFW concentrations are shown in Figure 3, and are consistent
with those produced in previous simulations as well as previous
studies of the MultiDark simulation \citep{PradaEtAl2012}.  As expected, halo
concentrations decrease with increasing halo mass. This
concentration-mass relationship is generally fit with a power-law model which
takes the form:  
\begin{equation}
c_{200} = \frac{c_{0}}{\left( 1+z \right)^{\beta}} \left( \frac{\mathrm{M}_{200}}{\mathrm{M}_{*}} \right)^{\alpha}
\end{equation}
where for our sample $z=0$, and we use M$_{*}= 10^{14} \mathrm{h}^{-1}
$M$_{\odot}$ (Figure 4). We first compute $r_{200}$ from the
cumulative density profile of each halo, and thus establishing  
M$_{200}$. Next, by fitting NFW profiles to these density profiles we
obtain concentration parameters. For MDR1 cluster halos, we find a
concentration-mass relation of  
\begin{equation}
c_{200} = \left( 4.775 \pm 0.022 \right) \left( \frac{\mathrm{M}_{200}}{10^{14} \mathrm{h}^{-1}
    \mathrm{M}_{\odot}} \right)^{-0.056 \pm 0.007}
\end{equation}

\begin{figure}
 \centering
 \includegraphics[scale=0.4]{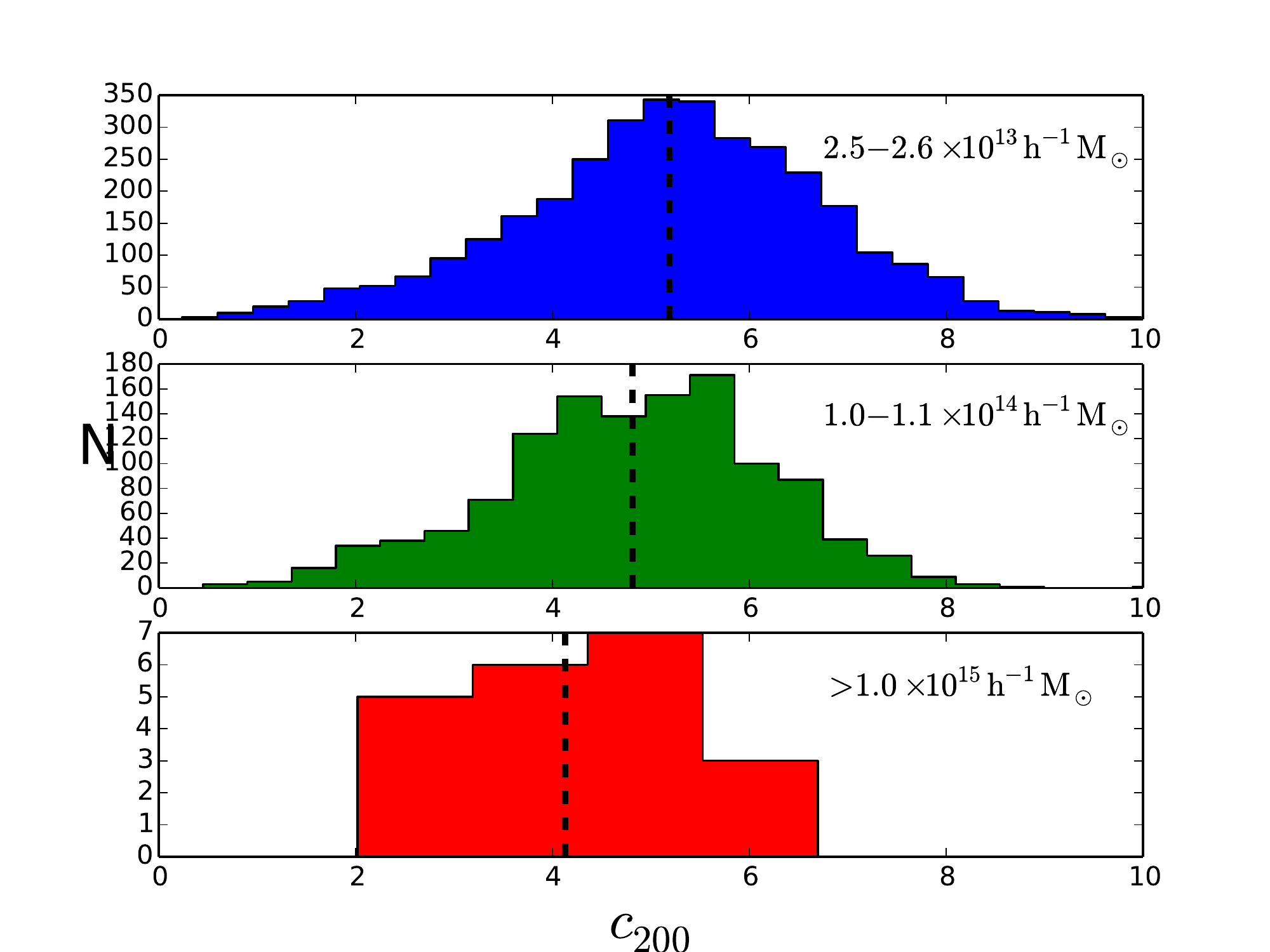}
\caption{Shown here are the distributions of intrinsic concentrations for all single virialized halos of each mass sample. From top to bottom: 1) Low Mass, 2) Medium Mass, and 3) High Mass.}
\end{figure}

From the intrinsic distributions of shape and concentration found for
each sample population, we find that solely due to line-of-sight
orientation, projected concentrations tend to be systematically higher
by about $20\%-50\%$ for virialized, single model halos. Halos with
intrinsically low concentrations have been found to suffer from a larger
orientation bias $\sim 50\%$, whereas higher intrinsic concentrations
tend to produce an over-concentration of $\sim 20\%$, albeit with much lower
scatter. Additionally, this trend tends to flatten out with
increasing distance from the cluster center. This means for a fixed 
halo concentration, inner regions of halos (those probed on or near strong
lensing scales) will bias higher than outer regions of halos (those probed
on weak lensing scales) if viewed along the major axis. This information is
captured for relaxed, low mass halos (Figure 5; See also
Table 3 for a complete summary). 

The underlying cause for this mismatch in over-concentration between
halos with intrinsically low and intrinsically high concentrations is
due to the magnitude of the change in shape as a function of radius. Low
concentration halos are shown to significantly change their shape
between $0.5 \cdot r_{200}$ and $r_{200}$, increasing in p and q,
$81\%$ and $80\%$ of the time with median differences of $\Delta p =
0.14$ and $\Delta q = 0.11$ (Figure 6). Halos with high intrinsic
concentrations increase in p and q $\sim 65 \%$ of the time, however,
the median value of the difference in axial ratio drops to $\sim 0.03$.

On the extreme end, we have shown that although a systematic bias
exists, certain low concentration halos ($c_{200}\sim 1-3$) with extreme
axis ratios can produce upwards of a factor of two higher in projected
concentration on a case-by-case basis. This can be seen most notably
in the low and medium mass samples at $r\sim 0.5 \cdot  r_{200}$.

Additionally, we also find that concentric ellipsoidal shells are well
described as being coaxial with one another, that is to say there
is insignificant amounts of twisting of isodensity surfaces for each mass
population. Alignment becomes even better with increasing radius,
where the misalignment is a maximum at the innermost radial
value. It should be noted, however, that alignment results are
expected to be biased low due to correlation between each ellipsoidal
surface and ones interior to it. Though the aggregate shows relatively
good alignment, it is again possible for individual cluster halos to produce
significant ($\la 30^{\circ}$) offsets between projected isodensities
located at strong and weak lensing scales. This fact alone complicates
things, however in the limit of large numbers of cluster halos, this
effect should be minimal in biasing reconstructed concentrations for
the population as a whole.  

\section{Summary and Conclusions}

We have shown that relaxed MultiDark MDR1 simulation (FOF) cluster
halos which are well described by a single NFW density profile are
primarily prolate spheroidal in geometry and become increasingly more spherical with
increasing radius from the cluster center of mass. Using shape on weak
and strong lensing scales as well as derived concentrations,
we analytically project these halos along the line of sight. In doing
so, we find that low mass clusters are typically over-concentrated by
about 56\% and 20\% at half $r_{200}$ for concentrations  between
1-3 and 7-10, respectively. At $r_{200}$ this enhancement drops to
38\% and 19\%. What this tells us is that the average projected
concentration differs by about 18\% for halos with intrinsically low
concentrations simply due to differences in halo geometry as a 
function of radius. Clusters which do not meet this criteria show the
opposite trend in shape, becoming more prolate with increasing radius. 

Strong lensing clusters are usually identified by their hard to miss
tangential or radial arcs, and are expected to represent a biased
population simply because large mass and alignment along the line of
sight are key ingredients in producing large Einstein radii. If these
lensing clusters are in fact preferentially aligned along the line of
sight (and are relaxed), we would expect that all else being 
equal weak lensing reconstructions should under-estimate the
concentration for a population of such objects.   
  
Projection effects aside, additional complications arise in measuring
halo concentration using strong and weak lensing. For example, halo
substructure can play a significant role in altering the shape of the
lens as seen by strong lensing \citep{MeneghettiEtAl2007},  along with
massive objects unassociated with the halo which lay along the line of
sight \citep{PuchweinHilbert2009}.  \citet{RedlichEtAl2012} also find
that cluster mergers bias high the distribution of Einstein Radii,
highlighting another source of potential bias. The weak lensing signal
can be diminished by things like atmospheric PSF, correlations in the
orientation of background galaxies due to large scale structure, among
the usual sources of uncertainty in measuring galaxy shapes (for a
review of galaxy shape measurement and correlation of galaxy shapes,
see \citealt{HoekstraJain2008}; for a discussion of cluster triaxiality
and projections of large scale structure see \citealt{BeckerKravtsov2011}).    

\section{Future Work}

An explicit prediction has been made regarding the discrepancy between
projected halo concentrations of cluster halos on characteristic
lensing scales. A natural next step would be to simulate the signal
produced from gravitational lensing by conducting mock weak and strong
lensing analyses. However, one would realistically need to include the
effects of baryons. Additionally, we plan to summarize the current
state of the field of galaxy cluster mass reconstructions in each of
the methods used. In future work, we will aggregate all measured NFW
mass/concentration pairs from these methods in order to shed light on
potential systematic observational biases, particularly on strong and
weak lensing scales.  

It has yet to be determined if this effect manifests itself in a
measurable way for the cluster halo population. Follow-up observations
of strong lensing clusters could be proposed as a way of testing the
veracity of this prediction. Knowing the intrinsic distribution of
cluster concentrations is difficult if not impossible due to the
degenerate nature of the reverse-projection process. However, the shape
of the distribution of measured concentrations due to lensing could
possibly contain hallmark characteristics which could indicate the
level of line of sight biasing of the population. With this
information known, ultimately a correction procedure could then be
proposed.

\section{Acknowledgements}
This work was primarily supported by NSF Grant 0908307. AMG would also
like to recognize Kristin Riebe for answering any questions he had
regarding the MultiDark database and colleagues Justin Bird and Markus
Rexroth for their time and feedback.

The MultiDark Database used in this paper and the web application
providing online access to it were constructed as part of the
activities of the German Astrophysical Virtual Observatory as result
of a collaboration between the Leibniz-Institute for Astrophysics
Potsdam (AIP) and the Spanish MultiDark Consolider Project
CSD2009-00064. The Bolshoi and MultiDark simulations were run on the
NASA's Pleiades supercomputer at the NASA Ames Research Center. The
MultiDark-Planck (MDPL) and the BigMD simulation suite have been
performed in the Supermuc supercomputer at LRZ using time granted by
PRACE. 

Many plots in this work were generated by astroML \citep{astroML}, a
python module for machine learning, data mining, and visualization of
astronomical datasets. 

\section*{Appendix}
The analytic form of the projected concentration is derived with the
assumption that the surface mass density remains a constant (that is, the
combination of shape and radial profile must change in such a way as to
keep the original 2-dimensional distribution the same). We start with
the scale convergence of a triaxial NFW halo as described in
\citet{SerenoEtAl2010a}:
\begin{equation}
\kappa_{s} = \frac{f_{geo}}{\sqrt{e_{P}}} \frac{\rho_{s}}{\Sigma_{cr}} r_{sP}
\end{equation}
where $e_{P}$, $r_{sP}$, and $f_{geo}$  are the projected inverse axis
ratio (the inverse of Equation 9), projected scale radius, and a
geometric elongation parameter, respectively. 
\begin{equation}
f_{geo} = \frac{e_{P}^{1/2}}{e_{\Delta}}
\end{equation}
The inverse projected axis ratio of a 3-dimensional ellipsoid viewed
at an arbitrary viewing angle has been worked out by
\citet{Binggeli1980} to be
\begin{equation}
e_{P} = \sqrt{\frac{j+l+\sqrt{(j-l)^{2}+4k^{2}}}{j+l-\sqrt{(j-l)^{2}+4k^{2}}}}
\end{equation}
where the intrinsic halo geometry (q is the semi-minor to semi-major
axis ratio; p is the semi-intermediate to semi-major axis ratio) and
viewing angle are input into: 

\begin{equation}
j = q^{2} \sin^{2}\theta + p^{2}\sin^{2}\phi \cos^{2} \theta + \cos^{2}\phi \cos^{2}\theta
\end{equation}
\[ k = p^{2} \cos^{2}\phi +\sin^{2}\phi \]
\[ l = \left(1-p^{2} \right) \sin\phi \cos\phi \cos\theta \]

The elongation parameter is defined in the following way:
\begin{equation}
e_{\Delta} = \left( \frac{e_{P}}{e_{1}e_{2}} \right)^{1/2} f^{3/4}
\end{equation}
\begin{equation}
f = e_{1}^{2} \sin^{2} \theta \sin^{2} \phi + e_{2}^{2}\sin^{2} \theta
\cos^{2} \phi + \cos^{2} \theta
\end{equation}
Remembering that the scale density $\rho_{s}$ is
\begin{equation}
\rho_{s} = \rho_{cr} \delta_{c}
\end{equation}
allows us to associate the extra geometric factor of $1/e_{\Delta}$ in
the scale convergence expression with the overdensity. 
\begin{equation}
\delta_{C} = \frac{1}{e_{\Delta}} \delta_{c}
\end{equation}
\begin{equation}
\frac{C^{3}}{\log{\left(1+C\right)} - \frac{C}{1+C}} = \frac{1}{e_{\Delta}}
\frac{c^{3}}{\log{\left(1+c\right)} - \frac{c}{1+c}}
\end{equation}
The extra geometric factor can be expressed in terms of the prolate 
spheroidal projected axis ratio $Q$, a function of the intrinsic axis
ratio q, and the angle between the major axis and the
line-of-sight. Above, we have conclude that the assumption of
prolateness is proven to be a reasonable one, thus the
surplus in concentration can be completely expressed in terms of
prolate spheroidal .
\begin{equation}
\frac{1}{e_{\Delta}} = \frac{Q^{2}}{q}
\end{equation}

\begin{table*}
\begin{center}
\caption{Concentration Enhancements}
\label{table-3}
\begin{tabular}{@{}lcccccc}
\hline
\hline
\textbf{Radial Scale}& & \textbf{Low Mass} & & \textbf{Medium Mass} \\ 
& $c_{200}$ & $\bar{\Delta}_{c_{200}} \pm \sigma_{\Delta}$ & $c_{200}$ & $\bar{\Delta}_{c_{200}} \pm \sigma_{\Delta}$ \\ 
\hline
$0.5 \cdot r_{200}$ & $[1-3]$ & $1.56 \pm 0.29$ & $[1-3]$ & $1.52 \pm 0.25$\\
& $[7-10]$ & $1.20 \pm 0.07$ & $[6-9]$ & $1.26 \pm 0.09$ \\
\hline
$ r_{200}$ & $[1-3]$ & $1.38 \pm 0.16$ & $[1-3]$ & $1.43 \pm 0.15$\\
& $[7-10]$ & $1.19 \pm 0.09$ & $[6-9]$ & $1.24 \pm 0.09$\\
\hline
$1.5 \cdot r_{200}$ & $[1-3]$ & $1.30 \pm 0.12$ & $[1-3]$ & $1.34 \pm 0.11$ \\
& $[7-10]$ & $1.17 \pm 0.07$ & $[6-9]$ & $1.22 \pm 0.09$ \\
\hline
$2 \cdot r_{200}$ & $[1-3]$ & $1.28 \pm 0.11$ & $[1-3]$ & $1.31 \pm 0.10$\\
& $[7-10]$ & $1.18 \pm 0.07$ & $[6-9]$ & $1.23 \pm 0.09 $ \\
\hline

\end{tabular}

\medskip
Average Concentration Enhancement ($\Delta_{c_{200}} \equiv
C_{200}/c_{200}$) on Weak and Strong Lensing Scales for Low and Medium
Mass Samples.
\end{center}
\end{table*}

\begin{figure}
\centering
\includegraphics[scale=0.4]{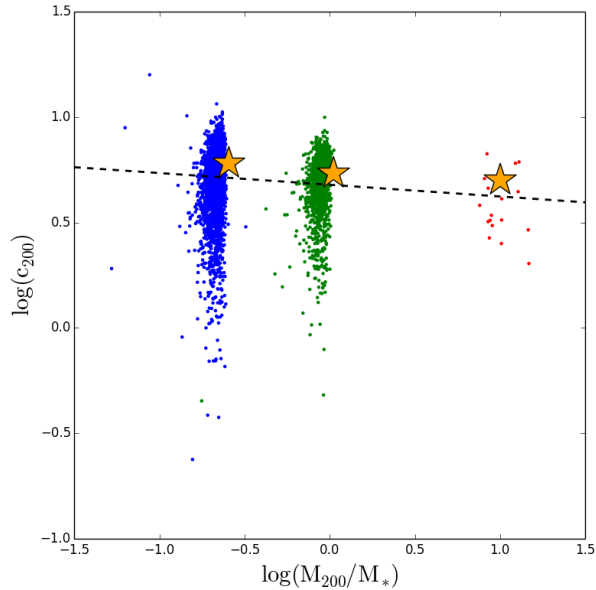}
\caption{Shown here is the intrinsic concentration-mass relationship for the MDR1 cosmological simulation run. Orange stars represent values of concentration from the analytical expression found in \citet{PradaEtAl2012} at redshift $z=0$ for values of halo mass corresponding to the sample used in this study.}
\end{figure}

\bibliography{paper1}

\begin{figure*}
\begin{center}$
\begin{array}{cc}
\includegraphics[scale=0.475]{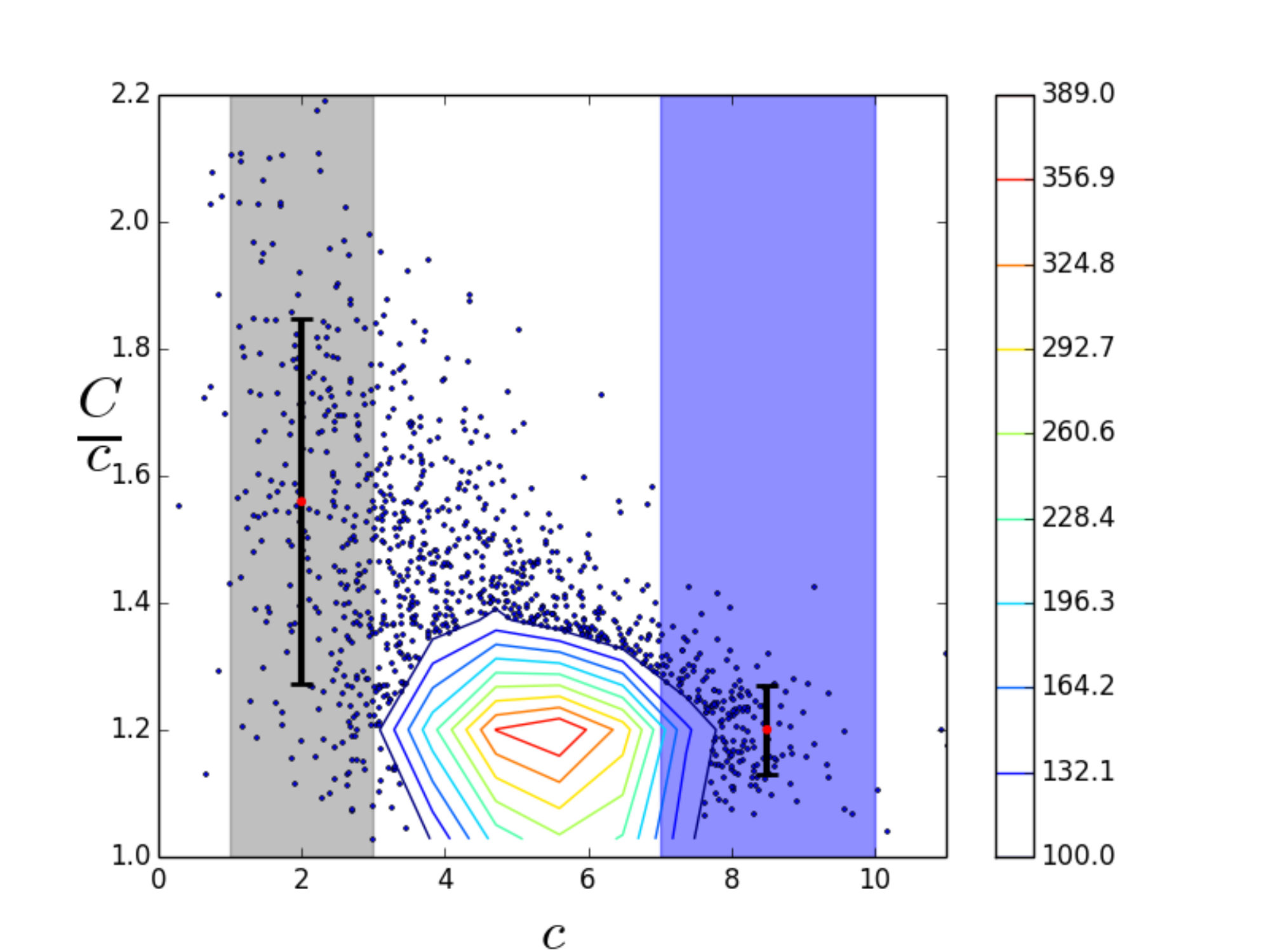} & \includegraphics[scale=0.475]{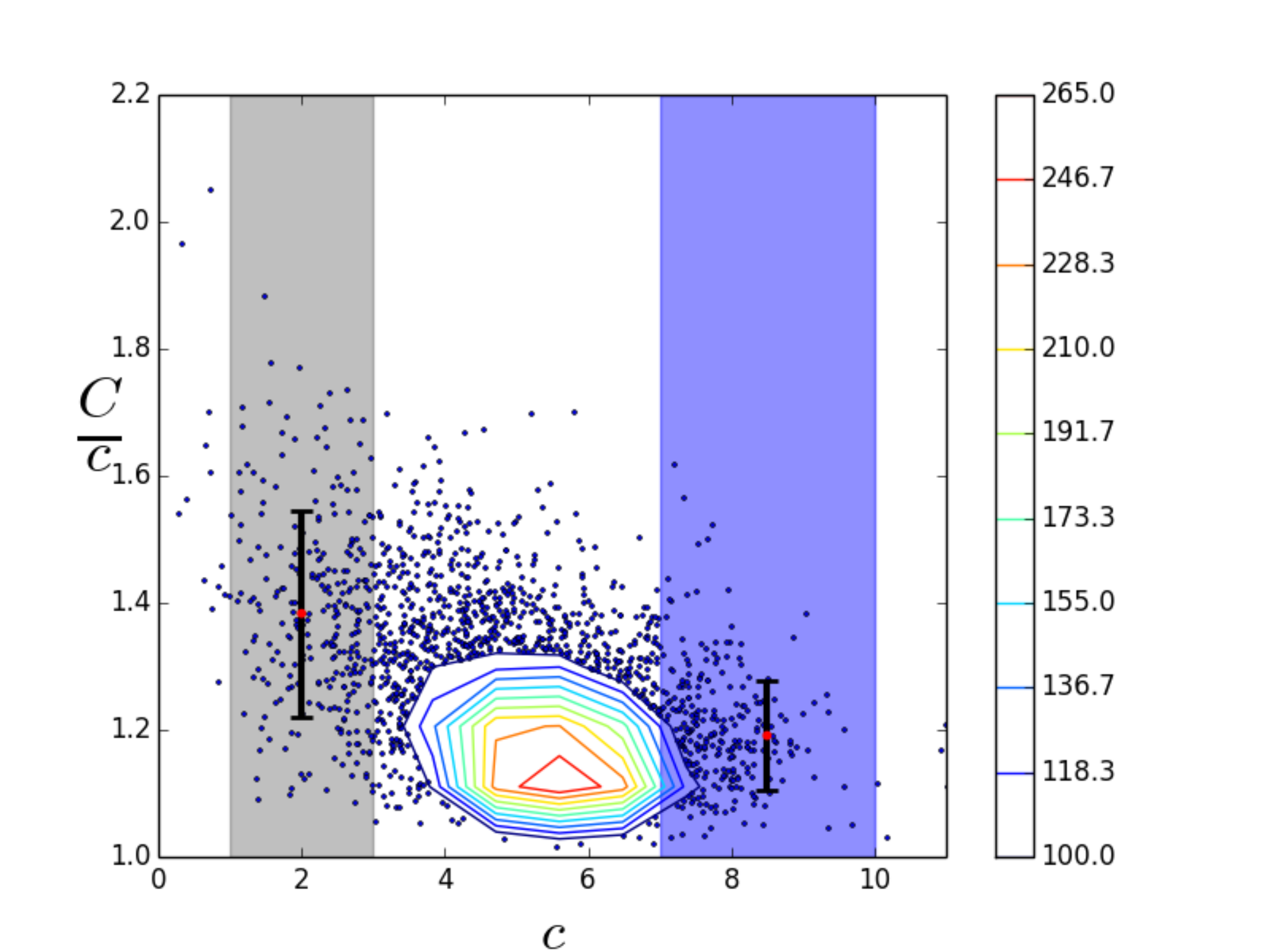}
\end{array}$
\end{center}
\caption{Shown here are concentration enhancements against intrinsic
  halo concentrations for line-of-sight oriented halos for the low
  mass, single-model, relaxed halo population.  In the left (right) panel,
  we show the population at $\sim 0.5 \cdot r_{200}$ ($\sim 1 \cdot
  r_{200}$). Overplotted are the mean and standard deviation of low
  ($1 \le c_{200} \le 3$) and high ($7 \le c_{200} \le 10$)
  concentration halos for comparison.} 
\end{figure*}

\begin{figure*}
\begin{center}$
\begin{array}{ccc}
\includegraphics[scale=0.3]{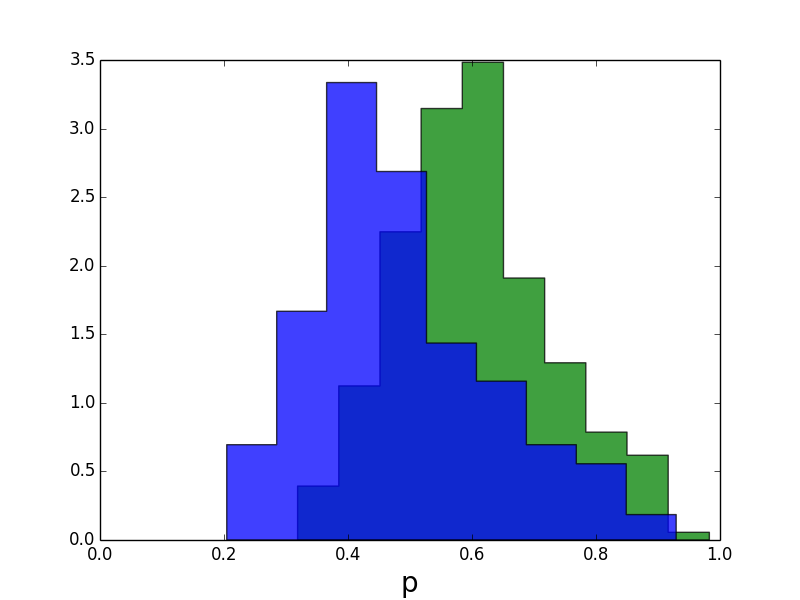} & \includegraphics[scale=0.3]{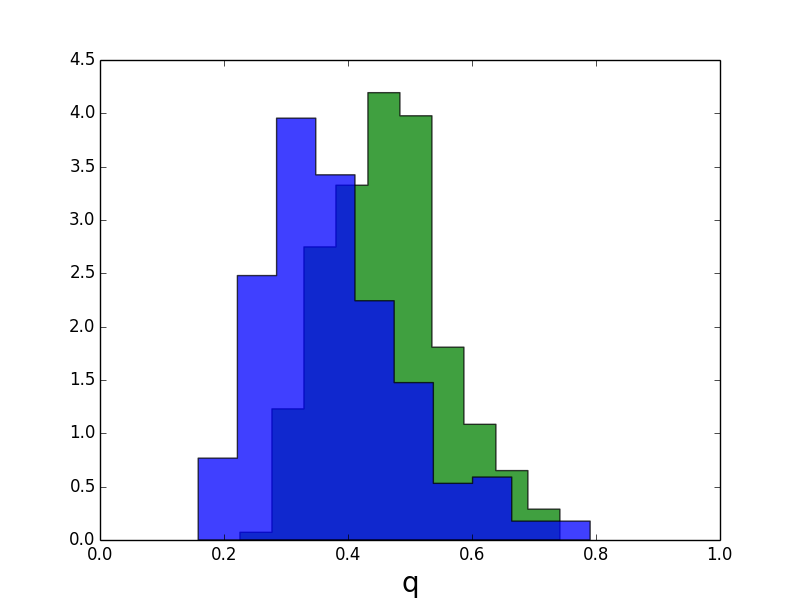} & \includegraphics[scale=0.3]{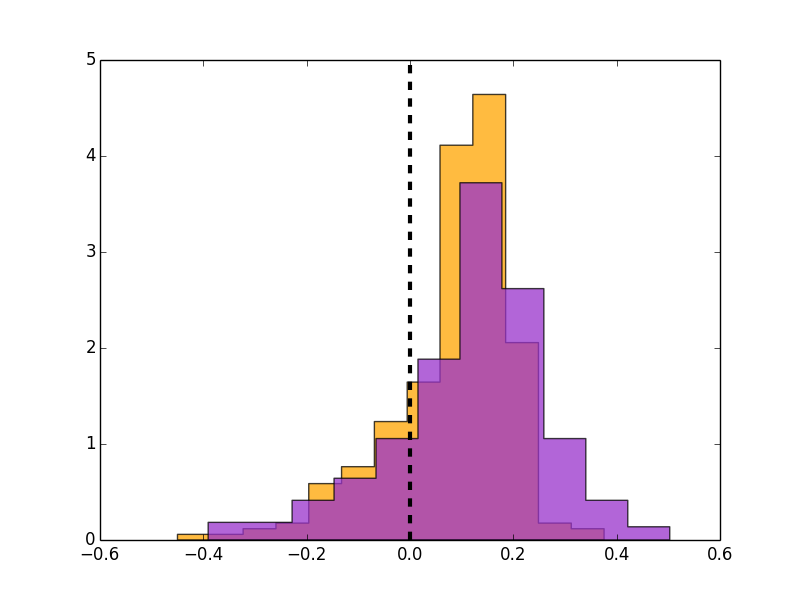} \\
\includegraphics[scale=0.3]{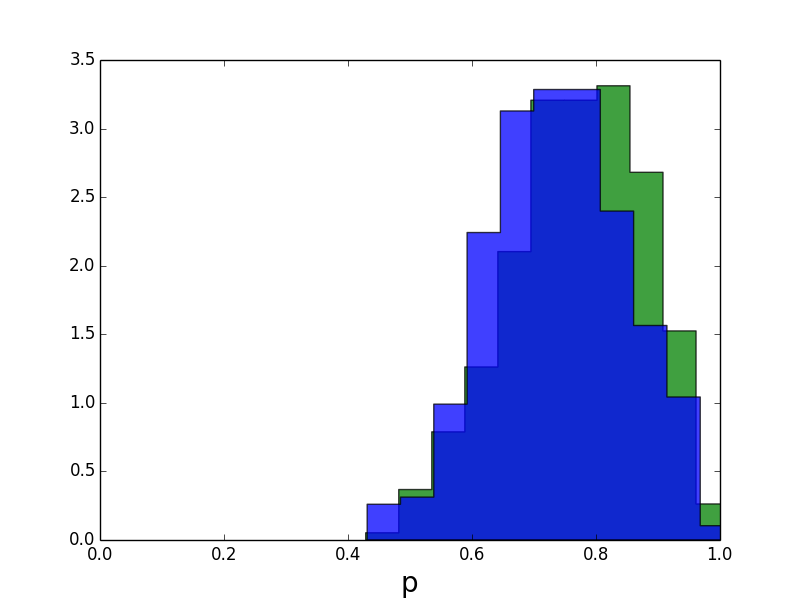} & \includegraphics[scale=0.3]{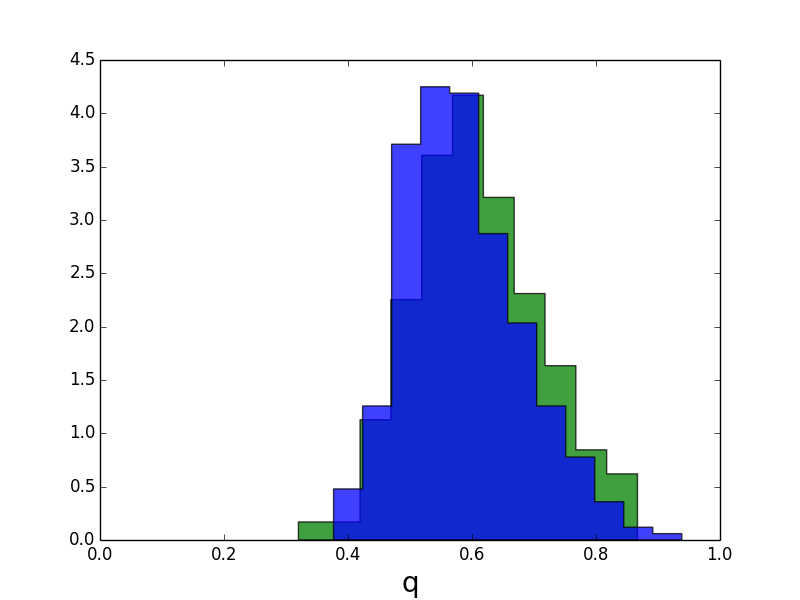} & \includegraphics[scale=0.3]{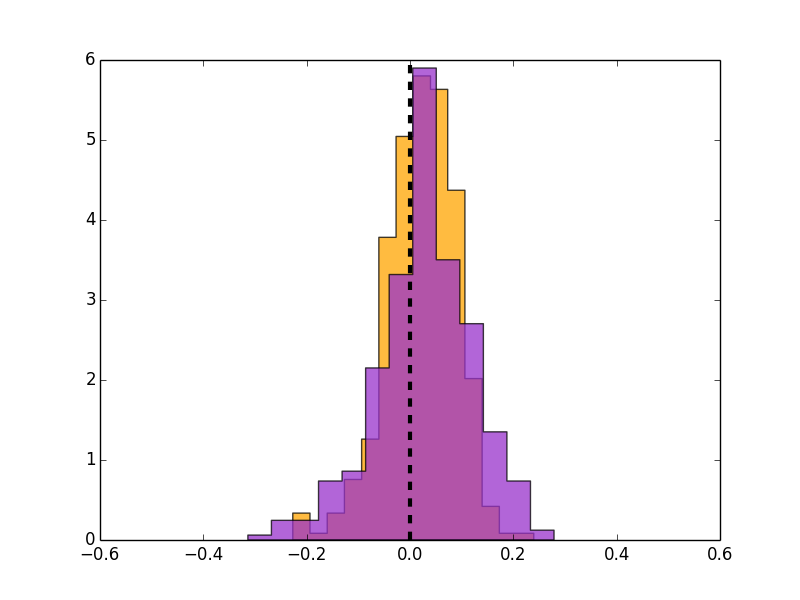}
\end{array}$
\end{center}
\caption{Halos with intrinsically low concentrations exhibit a much larger change in shape as a function of radius. \textit{Upper left and middle}: Normalized distributions of axial ratios p and q for low-mass, low-concentration (1-3) halos at half $r_{200}$ (blue) and $r_{200}$ (green). \textit{Bottom left and middle}: The distributions of axial ratios p and q for low-mass, high-concentration (7-10) halos at half $r_{200}$ and $r_{200}$. \textit{Upper right, (Lower right)}: Distributions of the differences in axial ratios p (purple) and q (orange) for low-concentration (high-concentration) low-mass cluster halos.}
\end{figure*}

\end{document}